\documentclass[runningheads]{llncs}
\usepackage{amsmath}
\usepackage{amsfonts}
\usepackage{graphicx}
\usepackage[caption=false]{subfig}
\usepackage{multirow}
\usepackage{physics}
\newcommand{\commented}[1]{}

\vspace{-2mm}
\begin{document}
\vspace{-2mm}
\title{Quantum-like Structure in Multidimensional Relevance Judgements}
%
%
\vspace{-2mm}
\author{Sagar Uprety\inst{1} \and
Prayag Tiwari\inst{2} \and
Shahram Dehdashti\inst{3} \and
Lauren Fell\inst{3} \and
Dawei Song\inst{1,4} \and
Peter Bruza\inst{3} \and
Massimo Melucci\inst{2}}
\authorrunning{S. Uprety et al.}
%
\institute{
The Open University, UK
\email{sagar.uprety@open.ac.uk}
\and
University of Padova, Italy
\email{} 
\and
Queensland University of Technology, Australia
\email{}
\and
Beijing Institute of Technology, China
\email{}
}
\maketitle              
\vspace{-8mm}
\begin{abstract}
A large number of studies in cognitive science have revealed that probabilistic outcomes of certain human decisions do not agree with the axioms of classical probability theory. The field of Quantum Cognition provides an alternative probabilistic model to explain such paradoxical findings. It posits that cognitive systems have an underlying quantum-like structure, especially in decision-making under uncertainty. In this paper, we hypothesise that relevance judgement, being a multidimensional, cognitive concept, can be used to probe the quantum-like structure for modelling users' cognitive states in information seeking. Extending from an experiment protocol inspired by the Stern-Gerlach experiment in Quantum Physics, we design a crowd-sourced user study to show violation of the Kolmogorovian probability axioms as a proof of the quantum-like structure, and provide a comparison between a quantum probabilistic model and a Bayesian model for predictions of relevance.

\end{abstract}
\vspace{-6mm}
\vspace{-2mm}
\section{Introduction}
\vspace{-2mm}
Relevance in Information Retrieval (IR) is widely accepted to be a cognitive feature, driving all our information interactions. All areas of research within IR thus strive to improve relevance of documents to a user's information need (IN). These research areas of IR can be broadly divided into two: system-oriented and user-oriented IR. Whereas the system-oriented viewpoint ties relevance to be an objective property of the document and query content, the user-oriented approach to IR views relevance as a cognitive property. Although IR fundamentally involves user interaction and decision-making, the user-oriented approach has been found harder to implement, especially in evaluating performance of IR systems. This is because of the variability in user judgements of relevance~\cite{cool_belkin_2011}. System-oriented IR thus sought to standardise IR evaluation, in which the user-cognitive notion of relevance was replaced by an objective, topical relevance. This led to evaluation methodologies based on the Cranfield and TREC type test collections. The user and all of his/her contexts were removed from the evaluation process. \\
\indent
Recent surge in availability of online user data has led to incorporation of more user context in the computation of relevance, e.g. in learning based ranking algorithms. This context is based on the user's past interactions with the system, in addition to user attributes like age, interests, etc. and current attributes like location, type of device, etc. The common feature in these various contexts is that they are static. They are determined before the point of user's interaction with the IR system. However, the process of IR is interactive and dynamic. In this paper, we focus on another type of context driving user interactions - dynamic context. Dynamic context is one which changes user's cognitive state \textit{during} information interaction. \\
\indent
One well-known example of when a dynamic context affects relevance is the phenomenon of Order Effect~\cite{order_Hogarth1992}. Order effects have been investigated and found to exist in IR in the presentation order of documents~\cite{Eisenberg1988_order,borlund_order,Huang2004_order,Xu2008_order}. For example, in a recent study reported in \cite{benyou_quantum_interf_Order}, two groups of participants were presented with a pair of documents $D_1$ and $D_2$ in two different orders. For some of such pairs, it was found that the relevance of a document judged by users is different depending on the order it was presented. Although the phenomenon may appear to have an intuitive explanation, it violates one of the fundamental assumptions of classical probability theory - joint distributions, where, for two random variables representing relevance of the documents - $R_1$ , $R_2$, $P(R_1, R_2) = P(R_2, R_1)$, i.e., the order of judging the documents does not matter. Order effects violate this fundamental assumption. Such order effects have also been investigated and reported in between the different dimensions of relevance, like Topicality, Understandability, Reliability, etc.~\cite{bruza_perceptions_of_document_relevance,Uprety:2018:IOE:3234944.3234972,uprety2019modelling}, where different orders of dimensions considered to judge a document lead to different relevance judgements.  \\
\indent
The field of Quantum Cognition~\cite{Busemeyer:2012:QMC:2385442} offers a generalised framework to model probabilistic outcomes of human decision-making. It has been successful in modelling and predicting order effects~\cite{Trueblood2011_quantum_account_ordereff,Wang2013} and other paradoxical findings where axioms of classical probability theory are violated~\cite{Pothos2009_quan_explan_irrational,Busemeyer2011_quantum_expl_prob_errors}. Conceptually, it challenges the notion that cognitive states have pre-defined values and that a measurement merely records them. Instead, the act of measurement creates a definite state out of an indefinite state and in doing so, changes the initial state of the cognitive system. In terms of relevance, we cannot pre-assign relevance of a document for a user. Instead, relevance is defined only at the point of interaction of the user's cognitive state with the document. Therefore, judgement of document $D_2$ first, changes user's initial state and the subsequent judgement of relevance of $D_1$ is different than when $D_1$ is judged before $D_2$. Should relevance of the documents for a user be a pre-defined entity, it would not be influenced by judgement of other documents and a joint distribution over relevance of the two documents would exist. We also say that these two measurements of relevance are incompatible with each other. That is, it is not possible to jointly consider the relevance of the two documents, at the same time. At the mathematical level, measurements in quantum theory are represented by operators, which in general, do not commute with each other.  \\
\indent
In a classical system, all measurements will commute with each other. However, conversely, commutativity of measurements does not necessarily imply that the system is  classical. Therefore, the type of measurements becomes imperative in identifying a quantum system. Even then, not all measurements on quantum systems generate data violating the classical probability theory. The system needs to be probed in a way which exploits the underlying quantum structure. In physics, this was done by experiments such as Stern-Gerlach and double-slit experiments~\cite{sakurai} which showed the violation of classical probability principles for microscopic particles like electrons and photons. In cognitive science too, several experiments performed by Tversky, Kahneman and colleagues showed such violations in human decision-making under uncertainty~\cite{Tversky1974}. 

Recently, an experiment protocol inspired by the Stern-Gerlach experiment in Physics has provided a new way to probe cognitive systems such that they exhibit a quantum-like structure~\cite{fell:dehdashti:bruza:moreira:2019}. By quantum-like structure we mean the representation of a system using the mathematical framework of quantum theory in order to model and predict the experimental data. In \cite{uprety2019modelling}, this experiment was performed in an IR scenario involving judgement of relevance with respect to different dimensions. Extending from the Stern-Gerlach protocol, in this paper we design a new experiment to show the violation of classical probability theory in multidimensional relevance judgements. We hypothesise that multidimensional relevance judgement has an underlying quantum-like structure, which when subject to appropriate measurement design can exhibit violations of classical probability theory. Specifically, we investigate the violation of a particular axiom of Kolmogorovian probability theory~\cite{kolmogorov_book}. Our results show that the experimental data indeed violates classical probability theory, and a quantum framework provides more accurate predictions to describe the data. This experiment not only shows the necessity of the quantum framework as an alternative for constructing probabilistic models, but also gives novel insights into user behaviour in IR. This understanding can contribute to improvement of interactive IR systems and we also discuss such implications in this paper.

\commented{In Section 2, we briefly describe the Stern-Gerlach experiment conducted in \cite{uprety2019modelling}, and the construction of a Hilbert space structure for modelling user's cognitive state. The details of our experiment are presented in Section 3, followed by the results in Section 4. In Section 5, we discuss the implications of our findings to IR. }

\commented{Recent references in multidimensional relevance \cite{shum2019multidimensional,palotti2018mm,su2018investigating,jiang2017comparing,albahem2019meta,chang2018relevance,haddad2018relevance,zhai2018matching}
}
\vspace{-4mm}
\section{Stern-Gerlach Inspired Protocol for Multidimensional Relevance}
\begin{figure*}[htp]
\vspace{-4mm}
\hspace{-20mm}
\subfloat[Asking three questions in TUR order]
{\label{figure-sg-cognitive-exp}
\vspace{0mm}
 \includegraphics[width=7.5cm]{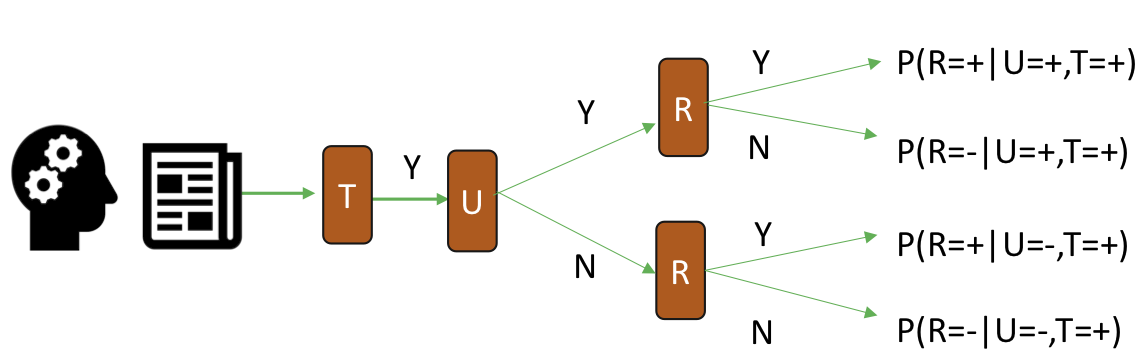}
 }
\subfloat[Asking three questions in TRU order]{\label{figure-sg-cognitive-exp-reverse}
\qquad
\includegraphics[width=7.7cm]{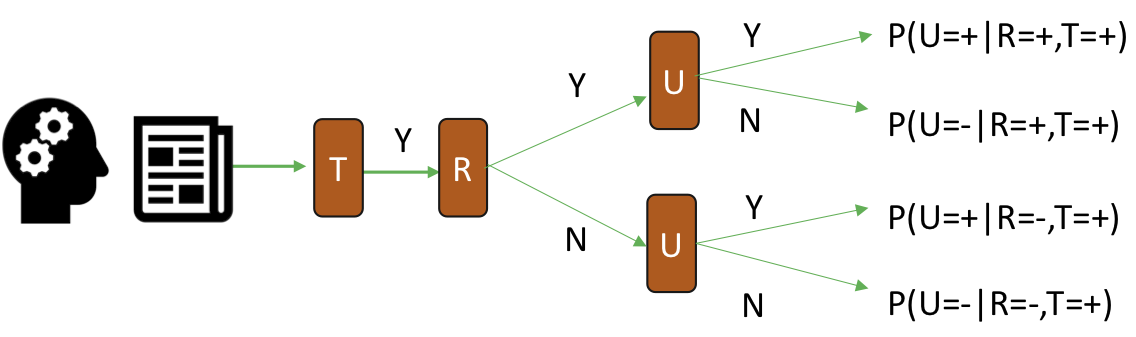}}
 \caption{S-G type Experiment to construct a complex-valued Hilbert Space}\label{sg-experiment}
 \vspace{-4mm}
\end{figure*}
\vspace{-2mm}
The basis of the research reported in this paper is the cognitive analogue of the Stern-Gerlach (S-G) experiment, originally conducted in \cite{uprety2019modelling}. The S-G experiment~\cite{sakurai} was an important milestone in quantum physics as it showed the non-classical behaviour of microscopic systems. The key was a particular design of the experiment which exploited the incompatibility between measurement of electron spin states along different axes. An electron has a particular property called spin, having two possible values - up (+), down (-), which can be measured along different axes. An electron may have spin state $+$ along the x-axis but state $-$ along y-axis. So the outcome of measurement of the spin property of the electron depends upon the axis of measurement. Also, any measurement of spin disturbs the system. If a measurement of spin is made along X axis and Z axis, then a third measurement along X axis may give a different answer than the first one. This phenomena is called measurement incompatibility, where two measurements cannot be jointly conducted on a system - one measurement disturbs the system and the other would then measure the changed system. \\
\indent
The S-G experiment also describes the minimum number of measurements required from a system to construct a complex-valued Hilbert Space structure. In particular, we need three incompatible measurements each with two mutually exclusive outcomes. We can use this arrangement of measuring properties of a quantum system to measure relevance of a document in IR. For this, we consider three dimensions of relevance: Topicality ($T$) - whether a document is topically relevant to a query, Understandability ($U$) - how easy it is to understand the content of the document, and Reliability ($R$) - how much can the document be relied upon. Each of these three dimensions can be posed as questions requiring a Yes/No type answer (denoted as $+$ and $-$ respectively) for a document. These three dimensions are important factors considered by users for deciding relevance. Besides, they are tied to a single document, unlike diversity or novelty, which is always considered in comparison with other documents. Certain dimensions like Interest, Habit, etc. are difficult to to ascertain via crowdsourcing. As reported in \cite{bruza_perceptions_of_document_relevance}, the different relevance dimensions can exhibit incompatibility for certain query-document pairs. \\
\indent
In \cite{uprety2019modelling}, three query-document pairs were designed in such a way as to potentially exhibit incompatibility between judgement of relevance with respect to different dimensions. The content of the documents was altered to introduce uncertainty in judging each of the three dimensions. The participants were presented with three questions related to three relevance dimensions, for each query-document pair, in line with the S-G design. Figure \ref{sg-experiment} shows the three questions asked to two different groups in different orders. More details about this design can be found in \cite{uprety2019modelling} and \cite{fell:dehdashti:bruza:moreira:2019}. This setup enables one to construct a complex-valued Hilbert space, which models the quantum-like structure of the user's cognitive state during information interaction.                                                                                                                                                                                                
\vspace{-4mm}
\subsection{Constructing Complex-Valued Hilbert Space}
\vspace{-2mm}
The first step in building a quantum probabilistic model is to construct a representation for the user's cognitive state. In the quantum framework, a complex-valued Hilbert space is used to represent a quantum system, and the state of the system is represented as a vector in this Hilbert space. \\
\indent 
Following the convention used in Quantum Physics, we represent any complex-valued vector $A$ in a finite dimensional Hilbert space as a ket vector $\ket{A}$ and its complex conjugate as a bra vector $\bra{A}$. The norm of this vector is the square root of its inner product with its conjugate - $|\braket{A}|^{1/2}$. For two such vectors, their projection onto each other is given as the square of their inner product - $|\bra{A}\ket{B}|^2$. Each vector is written as a linear combination of the vectors of the basis in which it is represented. For the purpose of representing the cognitive state of a person judging a document as topically relevant  or topically irrelevant, we consider a basis formed by two orthogonal vectors $\ket{T+}$ and $\ket{T-}$ respectively. Before a user considers a judgement of topicality, the cognitive state is indefinite with respect to considering the document as topically relevant or irrelevant. Both potentialities exist. We say that the cognitive state collapses to either $\ket{T+}$ or $\ket{T-}$ after the judgement. Before the judgement, we can represent the indefinite cognitive state in terms of probabilities of its potential responses. This is represented as a linear combination of the two basis states, weighted each by real or complex coefficients (called probability amplitudes), such that the square of the probability amplitude gives the probability of collapsing to the respective state. The initial state $S$ is thus written as:
\vspace{-2mm}
\begin{equation} \label{s-t}
\ket{S} = t\ket{T+} + \sqrt{1-t^2}\ket{T-}
\vspace{-1mm}
\end{equation}
\begin{figure*}[t]
\centering
      \includegraphics[width=\textwidth]{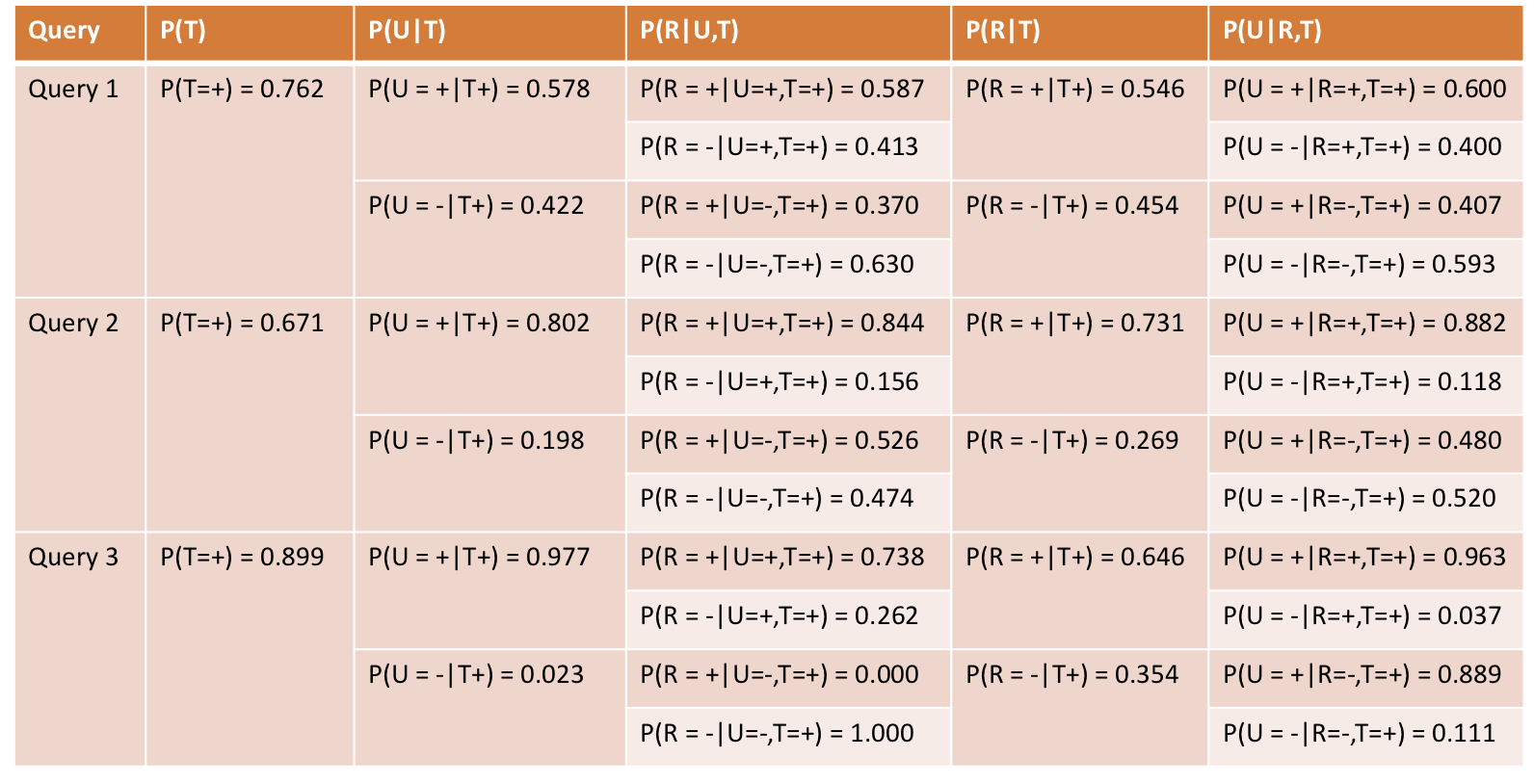}
\caption{Probabilities for the questions of TUR and TRU for the three queries}
\label{table-all_probabilities}
\vspace{-2mm}
\end{figure*}
In the S-G inspired experiment design, we ask the user sequential questions about judgement of Topicality (T), Understandability (U) and Reliability (R) in the order TUR or TRU, as shown in Figure \ref{sg-experiment}. Therefore we represent the cognitive state w.r.t Understandability and Reliability in term of Topicality:
\vspace{-2mm}
\begin{equation} \label{u-t}
    \ket{U+} = u\ket{T+} + \sqrt{1-u^2}\ket{T-}, \ket{U-} = \sqrt{1-u^2}\ket{T+} - u\ket{T-}
    \vspace{-2mm}
\end{equation}
$\ket{U-}$ is constructed using the fact that $\ket{U+}$ and $\ket{U-}$ are orthogonal. $u^2$ is the probability that users judge a document Understandable, given that they have judged it as Topically relevant.\\
\indent 
Refer to \cite{uprety2019modelling} (Section 3) or \cite{sakurai} (Chapter 1) for the necessity of using a complex-valued probability amplitude in the representation of Reliability in term of Topicality:
\vspace{-2mm}
\begin{equation} \label{r-t}
    \ket{R+} = r\ket{T+} + \sqrt{1-r^2}e^{i\theta_r}\ket{T-}, 
    \ket{R-} = \sqrt{1-r^2}e^{-i\theta_r}\ket{T+} - r\ket{T-}
    \vspace{-2mm}
\end{equation}
\indent 
The parameters ($u$, $r$ and $\theta_r$) comprise the construction of the Hilbert space for user's cognitive state w.r.t the interaction between the three dimensions. The parameter $t$ defines the initial state. The experiment design of Figure \ref{sg-experiment} was carried out in \cite{uprety2019modelling} for three queries. The results are listed in Figure \ref{table-all_probabilities}. 

\vspace{-5mm}
\section{Formulation of Research Hypotheses}
\vspace{-2mm}
Using the complex-valued Hilbert Space of multidimensional relevance, this paper aims to design an extended experiment to test the following research hypotheses: (1) Fundamental axioms of classical Kolmogorov probability are violated in a multidimensional relevance judgement scenario; (2) Probabilities obtained from the experiment can be better predicted with quantum than classical (Bayesian) probabilistic models. In the following two subsections, we mathematically formulate these hypotheses. 

\vspace{-3mm}
\subsection{Violation of Kolmogorov probability and Quantum Correction}
\vspace{-1mm}
 Quantum probabilities are generalisation of Kolmogorov probabilities. In fact, Kolmogorov probabilities are related to set theory which formalises Boolean logic. The following proposition gives one of their fundamental properties \cite{kolmogorov_book}:
 \vspace{-2mm}
 \begin{equation}\label{eq-kolmogorov}
     0=\delta=P(A \lor B)-P(A)-P(B)+P(A \land B)
        \vspace{-2mm}
 \end{equation}
 where $A$, $B$ are subsets of the set of all alternatives $\Omega$, and $P(A)$, $P(B)$ are the corresponding probabilities.  
The axiom will be violated if the value of $\delta$ is different from zero.\\ 
\indent 
In the quantum probability theory, the computation of probabilities are represented by projection operators for the events $U\pm$ and $R\pm$ corresponding to relevance or non-relevance with respect to Understandability and Reliability. The analogue of relation (\ref{eq-kolmogorov}) in quantum mechanics is given by the following definition \cite{vourdas2019probabilistic}:
\vspace{-2mm}
\begin{eqnarray}
\mathcal{D}(U\pm,R\pm)=\Pi(U\pm \lor R\pm)-\Pi(U\pm)-\Pi(R\pm)+\Pi(U\pm \land R\pm)
    \vspace{-2mm}
\end{eqnarray}
where projection operators $\Pi(U\pm)$ and $\Pi(R\pm)$ are given by:
\vspace{-2mm}
\begin{eqnarray}
\Pi(U\pm)=\ket{U\pm} \bra{U\pm}, \hspace{.5cm} \Pi(R\pm)=\ket{R\pm} \bra{R\pm}
\vspace{-2mm}
\end{eqnarray}
It is possible to prove that this
 quantum correction term $\mathcal{D}(U\pm,R\pm)$ is proportional to the commutator of the projection operators of $U\pm$ and $R\pm$~\cite{vourdas2019probabilistic} and can be thus obtained as :
\vspace{-2mm}
\begin{eqnarray} \label{eqn-quantum-delta}
\mathcal{D}(U\pm,R\pm)=\big[ \Pi(U\pm),\Pi(R\pm)\big]\left(\Pi(U\pm)-\Pi(R\pm)\right)^{-1}
\vspace{-2mm}
\end{eqnarray}
where $\left[A,B\right]$ stands for the commutator for two operators $A$ and $B$. The projection operator $\Pi(U+)$ is equal to the outer product of the state $\ket{U+}$ with itself, where the vector $\ket{U+}$ is computed using Equation \ref{u-t}. In order to construct the vector, first the Topicality basis is represented as the standard basis and hence the orthogonal vectors $\ket{T+}$ and $\ket{T-}$ are given as:
\vspace{-2mm}
\begin{equation} \label{eqn_standard_basis}
     \ket{T+} = \begin{pmatrix} 1 \\ 0 \end{pmatrix},
     \hspace{2mm}
     \ket{T-} = \begin{pmatrix} 0 \\ 1 \end{pmatrix}
\end{equation}
Thus, vectors $\ket{U+}$ and $\ket{U-}$ are given as:
\vspace{-2mm}
\begin{align}
         \ket{U+} = \begin{pmatrix} u \\ \sqrt{1-u^2} \end{pmatrix},
     \hspace{2mm}
     \ket{U-} = \begin{pmatrix} \sqrt{1-u^2} \\ -u \end{pmatrix}
\end{align}
Then the projector $\Pi(U+)$ is given as:
\vspace{-2mm}
\begin{equation} \label{eqn_U+_projector}
    \Pi(U+) = \ket{U+}\bra{U+} \\ \nonumber
    = \begin{pmatrix} u \\ \sqrt{1-u^2} \end{pmatrix} \begin{pmatrix} u & \hspace{2mm} \sqrt{1-u^2} \end{pmatrix} 
    = \begin{pmatrix} u^2 & \hspace{2mm} u\sqrt{1-u^2} \\ u\sqrt{1-u^2} & \hspace{2mm} 1-u^2 \end{pmatrix}
\end{equation}
Similarly,  $\Pi(R+)$ is :
\vspace{-2mm}
\begin{equation} \label{eqn_R+_projector}
    \Pi(R+) = \ket{R+}\bra{R+} \\ \nonumber
    = \begin{pmatrix} r \\ \sqrt{1-r^2}e^{i\theta_r} \end{pmatrix} \begin{pmatrix} r & \hspace{2mm} \sqrt{1-r^2}e^{-i\theta_r} \end{pmatrix} 
    = \begin{pmatrix} r^2 & \hspace{2mm} r\sqrt{1-r^2}e^{-i\theta_r} \\ r\sqrt{1-r^2}e^{i\theta_r} &  1-r^2 \end{pmatrix}
\end{equation}
From the values of $u, r$ and $\theta_r$ obtained in \cite{uprety2019modelling}, these projection operators can be constructed. The quantum analogue of $\delta$, can then be calculated from Equation (\ref{eqn-quantum-delta}). Value of $\delta$ obtained from our experiment is compared to that predicted by the classical (always zero) and quantum probability frameworks.

\vspace{-4mm}
\subsection{Quantum Probabilities vs Classical Probabilities}  
\vspace{-2mm}
The violation of Kolmogorovian probability axiom by a given system would likely lead to inaccurate predictions on the system using Kolmogorovian probability. This subsection formulates computation of conditional probabilities of relevance judgement along one dimension given another, using classical vs. quantum frameworks. They will be compared for our experimental data in Section 5.

For an initial state of the system $\ket{S}$, the probability of event $\ket{T+}$ in the quantum framework is given by $P(T+) = |\braket{T+}{S}|^2 = t^2$, i.e., square of projection of vector $\ket{S}$ onto vector $\ket{T+}$. The probability for sequence $U+$ following $T+$ is given as~\cite{Busemeyer:2012:QMC:2385442}:
\vspace{-2mm}
\begin{equation}\label{eqn-joint-prob}
P(U+,T+) = |\braket{U+}{T+}|^2|\braket{T+}{S}|^2
\vspace{-2mm}
\end{equation}
The quantum framework does not define joint probability of events $T$ and $U$, as in general $P(T+,U+) \neq P(U+,T+)$. As we can see $P(T+,U+) = |\braket{T+}{U+}|^2|\braket{U+}{S}|^2$, which for $\braket{U+}{S} \neq \braket{T+}{S}$ is not equal to $P(U+,T+)$ in Equation \ref{eqn-joint-prob}. The conditional probabilities are given according to Luder's rule~\cite{Busemeyer:2012:QMC:2385442,khrennikov2019basics} as:
\vspace{-4mm}
\begin{align}
        P_q(U+|T+) &= P(U+,T+)/P(T+|S) \\ \nonumber
                 &=\frac{|\braket{U+}{T+}|^2|\braket{T+}{S}|^2}{|\braket{T+}{S}|^2}\\ \nonumber
                 &= |\braket{U+}{T+}|^2 = u^2
\end{align}
Note that subscript $q$ is added to distinguish from classical conditional probability. Then $P_q(R+|U+,T+)$ is given as (see \cite{uprety2019modelling} Section 4.2 for derivation):
\vspace{-2mm}
\begin{align} \label{eqn-quantum-pred}
    P_q(R+|U+,T+) &= |\braket{R+}{U+}|^2 \\ \nonumber
                &= (ur)^2 + (1-u^2)(1-r^2) + 2ur\sqrt{(1-u^2)(1-r^2)}\cos{\theta_r}
\vspace{-2mm}
\end{align}
In contrast, classical probability theory has the basic assumption of commutativity of two events. Therefore the joint probability distribution always exists, which is the basis of calculating conditional probabilities in Bayes' rule. Consequently, for events $T$, $U$ and $R$ we have:
\vspace{-4mm}
\begin{equation} \label{eqn-joint-distr}
P(U+,R+,T+) = P(R+,U+,T+)
\vspace{-2mm}
\end{equation}
which can be written in terms of conditional probabilities as:
\vspace{-2mm}
\begin{equation}
   P(T+)P(R+|T+)P(U+|R+,T+) = P(T+)P(U+|T+)P(R+|U+,T+)
\vspace{-2mm}
\end{equation}
This enables calculation of conditional probabilities using the Bayes rule:
\vspace{-2mm}
\begin{align} \label{eqn_bayesian_pred}
  P(R+|U+,T+) &=   \frac{P(U+|R+,T+)P(R+|T+)}{P(U+|T+)} 
\end{align}
Similarly, the other conditional probabilities can be obtained. Again, note that the probabilities in Equations (\ref{eqn_bayesian_pred}) and (\ref{eqn-quantum-pred}) are different because of the difference in the underlying assumption of commutativity or joint probability.

\vspace{-2mm}
\vspace{-4mm}
\section{Experiment}
\vspace{-4mm}
\subsection{Methodology}
\begin{figure}[t]
\centering
\includegraphics[width=0.8\textwidth]{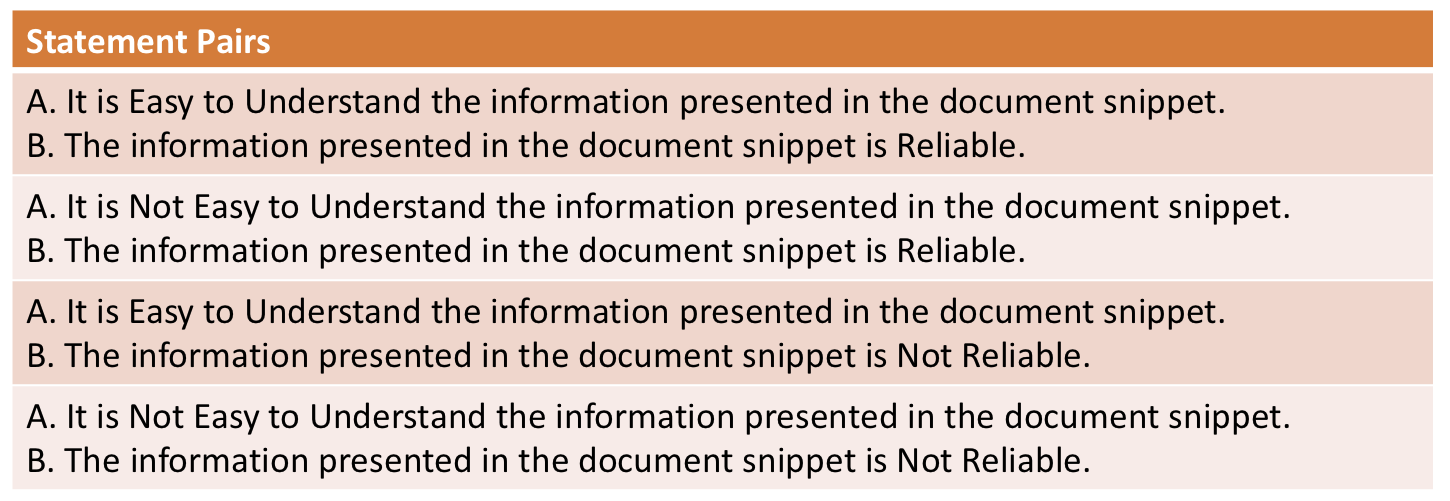} 
\caption{Four pairs of statements for conjunction and disjunction questions}
\label{figure-statement-pairs}
 \vspace{-6mm}
\end{figure}

\begin{figure*}[b]
\hspace{-20mm}
\subfloat[Design for disjunction question]{\label{figure-disjunction-question}
\vspace{0mm}
\includegraphics[width=7.5cm]{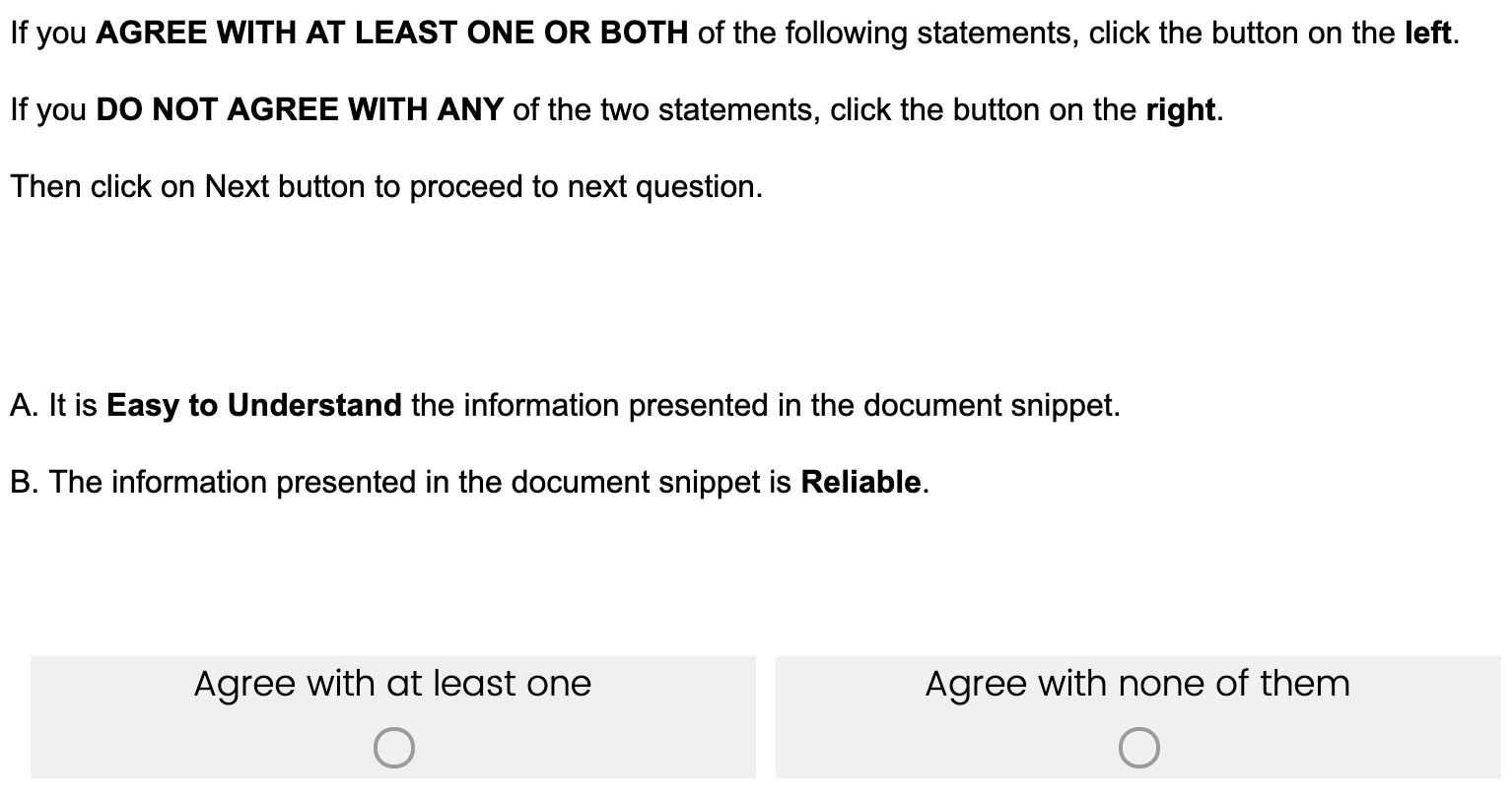}}
\subfloat[Design for conjunction question]{
\label{figure-conjunction-question}
\qquad
\vspace{-5mm} \includegraphics[width=7.7cm]{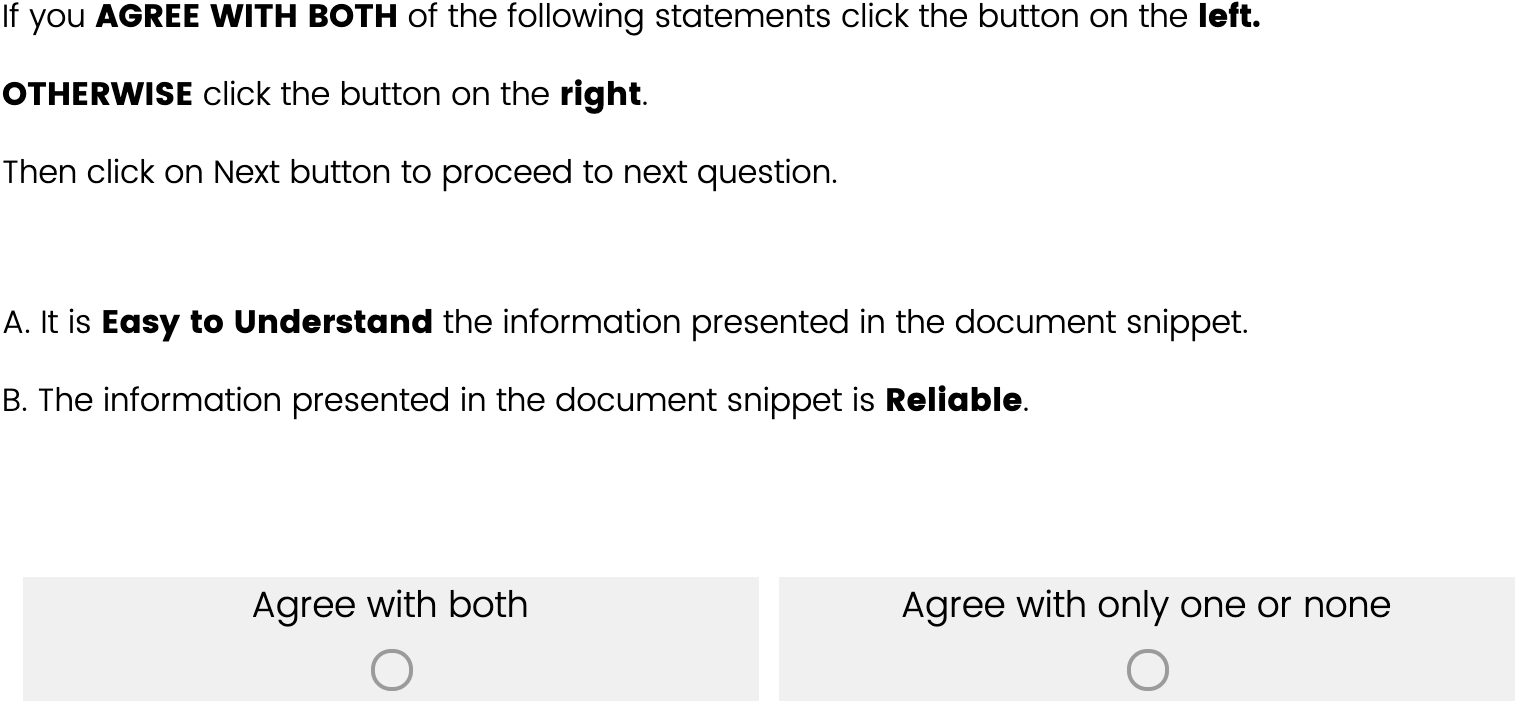}}
\caption{Conjunction/Disjunction question design}
\label{conjunction-disjunction-questions}
\end{figure*}

\vspace{-2mm}
The main aim of this experiment is to investigate the violation of Equation \ref{eq-kolmogorov}. We already have the single question probabilities from the experiment in \cite{uprety2019modelling} and we need to obtain the probabilities of conjunction and disjunction. We do so by posing questions about Understandability and Reliability at the same time, as a pair, rather than sequentially. Each of the dimensions have two outcomes (e.g. Reliable or Not Reliable) and therefore we construct four pairs of statements, as listed in Figure \ref{figure-statement-pairs}. For the disjunction measurement, we ask the participants to select whether they agree with at least one of the two statements or none of them (corresponding to a Boolean Or condition). For a conjunction measurement on each of the four statement pairs, we ask the participants whether they agree with both of the questions or not. Figures \ref{figure-disjunction-question} and \ref{figure-conjunction-question} show the designs for the disjunction and conjunction questions for a query-document pair. We now have a total of eight such questions and we follow a between-subjects design such that a participant is shown only one of these eight questions randomly. Note that we are able to use the probabilities from the experiment in \cite{uprety2019modelling} because our experiment is a between-subjects design. The same participant is not asked all the questions - to avoid memory bias. The design is summarised in the following steps for each of the three query-document pairs:
\vspace{-2mm}
\begin{enumerate}
\item The participants are shown information need, query and document snippet.
\item Next, they are asked a Yes/No question about the Topicality of the document. This is to prepare the cognitive state of all participants by projecting their initial/background state onto the Topicality subspace of the underlying Hilbert space constructed in the previous experiment in \cite{uprety2019modelling}.
\item Lastly, they are randomly shown one of the eight possible conjunction or disjunction questions and asked to choose the appropriate answer. 
\end{enumerate}

\vspace{-0mm}
\begin{figure*}[htp]
\hspace{-20mm}
\subfloat[Design for disjunction question]
{\label{figure-disjunction}
\vspace{-4mm}
\hspace{10mm}
\centering
 \includegraphics[width=0.5\textwidth]{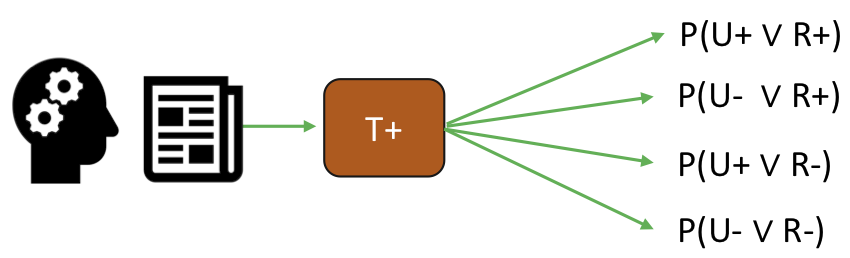}
 }
\subfloat[Design for conjunction question]
{\label{figure-conjunction}
\vspace{-4mm}
\qquad
\includegraphics[width=0.5\textwidth]{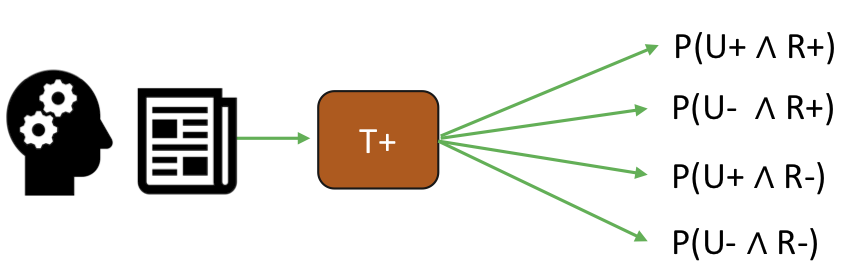}}
\vspace{-2mm}
 \caption{Experiment design}\label{new-experiment}
 \vspace{-4mm}
\end{figure*}
\vspace{-4mm}

\vspace{-4mm}
\subsection{Participants and Material}
\vspace{-2mm}
We recruited 335 participants for the experiment using the online crowd-sourcing platform Prolific (prolific.ac). The study was designed using the survey platform Qualtrics (qualtrics.com/uk). The participants were paid at a rate of \textsterling 6.30 per hour. We sought the participants' consent and complied with the local data protection guidelines. The study was approved by The Open University UK's Human Research Ethics Committee with reference number HREC/3063/Uprety.
\indent 
We use the same set of three query-document pairs for our experiment as used in \cite{uprety2019modelling}, as we have reused some of their data. Each participant was shown the three queries (and the documents) and were asked to judge the topicality of the document and one of the eight questions (so we obtain probabilities like $P(U+ \lor R+|T+)$, etc.) Thus the participants can be said to be divided into eight groups for a between-subjects design.
\vspace{-4mm}
\section{Results and Discussion}
\vspace{-3mm}
\subsection{Violation of Kolmogorov probability axiom}
\begin{figure}[h!]
\vspace{-3mm}
\centering
\hspace*{-0mm}\includegraphics[width=0.6\textwidth]{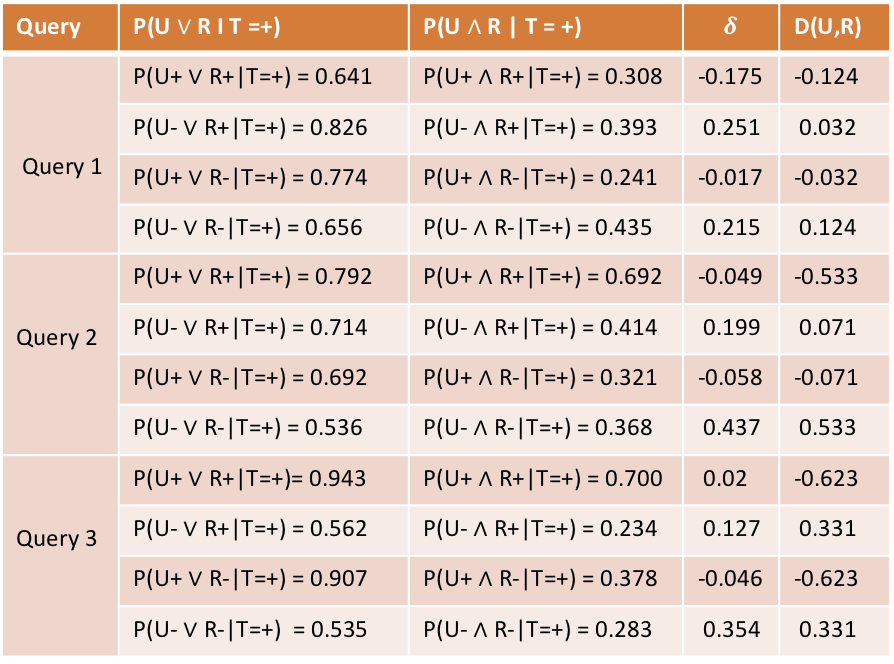}
\vspace{-4mm}
\caption{Probabilities for conjunction and disjunction questions and associated violation from Kolmogorovian probability}
\vspace{-4mm}
\label{table-deltas}
\vspace{-4mm}
\end{figure}

\vspace{-2mm}
The probabilities of conjunction and disjunction of the Understandability and Reliability questions are reported in Figure \ref{table-deltas}. In order to compute the $\delta$ reported in Equation \ref{eq-kolmogorov}, we also need the two probabilities related to single questions $U+$ and $R+$, apart from the conjunction and disjunction probabilities. These single question probabilities are obtained from the results in \cite{uprety2019modelling} (listed in Figure \ref{table-all_probabilities}). Then, we calculate $\delta = P(U\pm \lor R\pm|T+) + P(U\pm \land R\pm|T+) - P(R+|T+) - P(U+|T+)$. In Figure \ref{table-deltas} we see that $\delta$ is  different from zero for all the three queries, although according to classical probability we expect that $\delta$ would be zero in all cases. Equation (\ref{eqn-quantum-delta}), based on the projection operators in quantum probability, gives predictions of $\delta$, as are shown in the last column of the table. \\
\indent 
The violation of classical probability is a result of non-commutative structure of operators for $U$ and $R$. As we can see, if operators of $U$ and $R$ commute with each other, the quantum correction term in the Equation (\ref{eqn-quantum-delta}) approaches zero (the commutator is zero). In fact, the probability values obtained may violate some of the other basic axioms of classical/Kolmogorovian probability. For example, for Query 2, we can see that $P(U- \land R+|T+) = 0.414$ and $P(U-|T+) = 0.198$ which clearly violates $P(A,B) < P(A)$. Also, for this query, $P(U- \land R-|T+)$ is greater than both $P(U-|T+)$ and $P(R-|T+)$. This type of violation has been termed as conjunction fallacy in the cognitive science literature~\cite{Tversky1983_conjunction}. Quantum models have been previously used to explain such violation~\cite{Busemeyer2011_quantum_expl_prob_errors} where the fundamental notion of incompatibility in judgements is identified as the potential cause.

\vspace{-4mm}
\subsection{Comparison of Quantum and Classical probability predictions}
\begin{figure*}[htp]
\vspace{-4mm}
   \subfloat[]{\label{figure-q1-qp-cp}
      \hspace{-10mm}\includegraphics[width=0.60\textwidth]{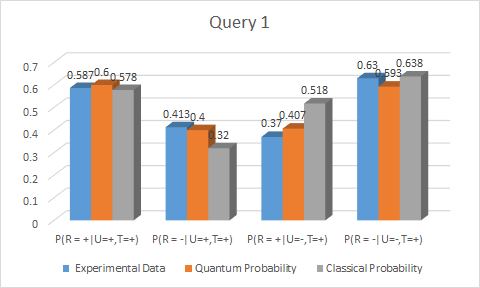}}
   \subfloat[]{\label{table-q2-qp-cp}
      \includegraphics[width=0.60\textwidth]{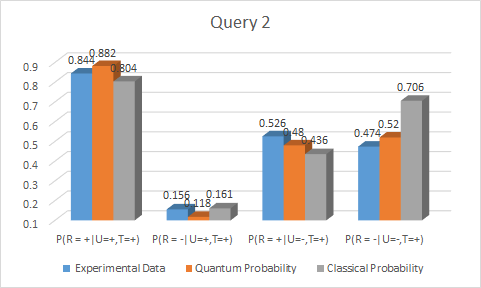}}
      \vspace{-2mm}
 \caption{How Quantum and Classical Probabilities compare with the experimental data for Query 1 and Query 2}\label{figure-qp-cp}
 \vspace{-10mm}
\end{figure*}

\vspace{-2mm}
Figure \ref{figure-qp-cp} shows a comparison between quantum and classical probabilities with the experimental data for first two queries. The data for Query 3 had many probabilities close to 0 (see Figure \ref{table-all_probabilities}) and hence the sample became too small for a meaningful comparison. The probabilities are calculated for prediction of judgement of Reliability given the participant has judged Understandability and Topicality (positively), using equations derived in Section 2.3. Bayesian probabilities, in some cases, are significantly different from experimental data ($P(R+|U-,T+)$ for query 1 and $P(R-|U-,T+)$ for query 2). Quantum probabilities are consistently closer to the experimental data.

The Bayesian probabilities, as mentioned earlier, are based on the chain rule $P(R+,U+,T+) = P(R+|U+,T+)P(U+|T+)P(T+)$. The fundamental assumption here is that the variables corresponding to $R$, $U$ and $T$ can be jointly measured. In terms of the judgement process, this implies that a user can jointly consider information regarding the Reliability, Understandability and Topicality of a document with respect to the query. The incompatibility revealed in \cite{uprety2019modelling} and the order effects shown in \cite{bruza_perceptions_of_document_relevance} suggest that this is not always the case in general. Therefore we see Bayesian predictions deviate from the experimental data. As the quantum probability theory based on the Hilbert space model is free from this assumption of compatibility, it provides a promising alternative model that gives predictions closer to the experimental data. In fact, the modelling of incompatibility of different judgement perspectives forms one of the pillars of the Quantum Cognition research framework. 
\vspace{-2mm}
\vspace{-3mm}
\section{Implications for IR}
\vspace{-3mm}
Quantum models can capture richer cognitive interactions, by way of generalising some of the constraints of classical models like commutativity. Here we discuss a few cases where our findings can inform the design of IR systems and algorithms.\\
\indent
The impossibility of jointly modelling Reliability and Understandability (which leads to the Kolmogorovian axiom violations) can be attributed to the fact that humans make decisions in a sequential manner and consideration of one dimension affects the judgement of the next dimension. Therefore, different orders of consideration of dimensions would lead to different final relevance judgements,\textit{ making the order a factor in the variability of relevance judgements by users}. When using an IR system to perform a task or make an important decision, there might be a particular order of dimensions which can lead the user to make an optimal decision. For example, for a health related query, a user might find a document difficult to understand, which may affect his or her judgement of Reliability and hence the overall relevance. However, if another user first judges reliability and finds it highly reliable, the judgement of understandability might be different. The IR system can help users to consider the optimum sequence of dimensions and thus maximise the utility, by providing extra information. For example, if the system can also provide information about the Reliability of the document in terms of a Reliability score or ratings by other users, it can reduce uncertainty in judgement and thus minimise the influence of judgement of other dimensions. Thus, for the given medical document, the low understandability might not affect the perception of Reliability. \\
\indent
Secondly, quantum probabilistic models can replace Bayesian models used in IR algorithms for ranking and evaluation. For example, in \cite{Palotti_multidimensional_evaluation}, a multidimensional evaluation metric is proposed where the gain provided by a document is written as a function of the joint probability of relevance with respect to different dimensions, e.g. $P(T,U,R,...)$. Similar assumptions have also been made in \cite{Palotti_understandability,zuccon_understandability_evaluation}. For documents exhibiting incompatibility between different dimensions, predictions from such a model will be inaccurate. A probabilistic model based on non-commutative operator algebra, accounting for the incompatibility between different dimensions, needs to be considered. \\
\indent 
Finally, these results of violation of classical probability theory calls for further user behaviour experiments to be conducted in IR that further exploit the Quantum-like Structure in human judgements. It would require novel experimental protocols like that of Stern-Gerlach, Double-slit experiment, etc., to generate data beyond the modelling capacity of classical probability theory. Such experiments in themselves might lead us to new insights into user behaviour in IR and information based decision-making in general.
\vspace{-4mm}
\section{Conclusion}
\vspace{-2mm}
Extending a quantum-inspired experiment protocol, in this work, we begin with the hypothesis that the multidimensional property of relevance has an underlying quantum cognitive structure which can be shown as violation of certain classical (Kolmogorovian) probability axioms. A particular experimental design is reported which can exploit the quantum cognitive structure. The data shows violation of one of Kolmogorovian probability axioms. We further show that quantum probability theory is a better alternative to model multidimensional relevance judgements than its classical counterpart, i.e. Bayesian model. Finally, we highlight important implications of our research findings to the design of IR algorithms system and user experiments. 
\vspace{-4mm}
\section*{Acknowledgements}
\vspace{-2mm}
Authors affiliated to the universities in UK, Italy and China are funded by the European Union's Horizon 2020 research and innovation programme under the Marie Sklodowska-Curie grant agreement No 721321, National Key Research and Development Program of China (grant No. 2018YFC0831704) and Natural Science Foundation of China (grant No. U1636203). Authors affiliated to QUT, Australia are supported by the Asian Office of Aerospace Research and Development (AOARD) grant: FA2386-17-1-4016.

%
%
%
%
\bibliographystyle{splncs04}
\bibliography{bibliography}
\end{document}